\documentclass[aps,prl,groupedaddress,twocolumn,superscriptaddress]{revtex4}

\usepackage{bbm}
\usepackage{mathrsfs}
\usepackage{amsmath}
\usepackage{amsfonts}
\usepackage[colorlinks=true,citecolor=blue,anchorcolor=blue]{hyperref}
\usepackage{graphicx,epstopdf}
\usepackage{subfigure}
\usepackage{epsfig}
\usepackage{dcolumn}
\usepackage{bm}
\usepackage{color}
\usepackage{natbib}
\usepackage{amssymb}
\usepackage{xcolor}
\usepackage{braket}
\usepackage{hyperref}

\begin{document}
\title{Long-distance strong coupling of magnon and photon: Effect of multi-mode waveguide}
\author{Yang Xiao}
\affiliation{Department of Physics, Nanjing University of Aeronautics and Astronautics, Nanjing 210016, China}

\begin{abstract}
\textbf{Coupled mode theory predicts that the long-distance coupling between two distant harmonic oscillators is in the weak coupling regime. However, a recent experimental measurement observed strong coupling of magnon and critically-driven photon with a distance of over two meters. To explain the discrepancy between theory and experiment, we study long-distance coupling of magnon and photon mediated by a multi-mode waveguide. Our results show that strong coupling is achieved only when both critical coupling and multi-mode waveguide are involved. The former reduces the damping while the latter enhances the coupling strength by increasing the pathways of coupling magnon and photon. Our theory and results pave the way for understanding the long-distance coherence and designing the magnon-based distributed quantum networks.}
\end{abstract}

%\pacs{
%72.25.Mk,  % Theory of electronic transport; scattering mechanisms
%73.22.-f,  % Molecular electronic devices
%73.40.Gk,   % transport in nanoscale systems
%}

\maketitle

\section{I. Introduction}

Strong coupling of magnon and photon~\cite{soykal2010prl,huebl2013prl,tabuchi2014prl,goryachev2014prapplied,zhang2014prl,bai2015prl,wang2018prl,cao2015prb,boventer2020prr,harder2018prl,yu2019prl,bhoi2019prb,grigoryan2018prb,yuan2022pr,li2018prl,yuan2020prl,sun2021prl,yang2024prapplied} is important for the understanding of magnon-photon coupling mechanism~\cite{lachance2019hybrid} and for potential applications of magnon-based quantum transduction~\cite{tabuchi2015coherent,lachance2020entanglement}. Especially, long-distance strong coupling is pivotal for the realization of magnon-based quantum network and quantum communication~\cite{rameshti2022pr}.

In general, magnon-photon coupling can be realized by placing a yttrium iron garnet (YIG) sphere at the antinodes of a microwave cavity~\cite{soykal2010prl,huebl2013prl,tabuchi2014prl,goryachev2014prapplied,zhang2014prl,bai2015prl,wang2018prl,cao2015prb}. Under the external magnetic field, the precessing magnetizations (magnon) of YIG interact with the microwave magnetic field through magnetic dipolar interaction. Due to the coherent nature of interaction, the coupling is usually called coherent magnon-photon coupling. Currently, many studies in this field belong to coherent coupling. Contrary to coherent coupling, another type of coupling can be implemented by putting the YIG sphere at the nodes of a cavity~\cite{harder2018prl}. In this case, the magnetic dipolar interaction and thus coherent coupling are very small. Therefore, the traveling photons, instead of cavity photons, inside the cavity induce an indirect magnon-photon coupling by interacting magnon and cavity photon with common traveling photons. This coupling is called dissipative coupling. To completely eliminate the magnetic dipolar interaction, the waveguide, instead of cavity, is usually used. In a very recent experiment~\cite{yang2024prl}, a YIG sphere and a dielectric cavity resonator are placed on a microstrip line with a long-distance separation of up to about two meters. Experimental observations in the linear response regime show anomalous strong coupling which has not been theoretically explained yet. Moreover, long-distance coupling is also observed by introducing the gain into the the magnon-photon coupling with a nonlinear response~\cite{rao2023prl}.

It is generally believed that the long-distance coupling is weak based on the coupled-mode theory~\cite{Manolatou1999ieee}. This is because the coupling is mediated by the energy dissipation to the waveguide. Therefore, the cooperativity will be always smaller than one and the system is in the weak coupling regime~\cite{yang2023prb2,Karg2019pra}. In comparison to experimental observations of strong coupling~\cite{yang2024prl}, one conclude that new physical mechanisms have not been revealed yet. In this work, we find that, as the waveguide consists of multiple propagation modes, instead of a single propagation mode, strong coupling can be achieved.

The remainder of this paper is organized as follows. In Sec. II, we review the coupled-mode theory of single-mode waveguide and give the results of waveguide/cavity and magnon-photon coupling. In Sec. III, we introduce two propagation modes in the waveguide and derive the transmission coefficient. The results of multi-mode waveguide are analyzed and compared to single-mode waveguide. The possible origin and detection method are discussed for multiple modes in the waveguide. Conclusion is presented in Sec. IV.

\begin{figure}[t]
\centering
\includegraphics[angle=0,scale=0.6]{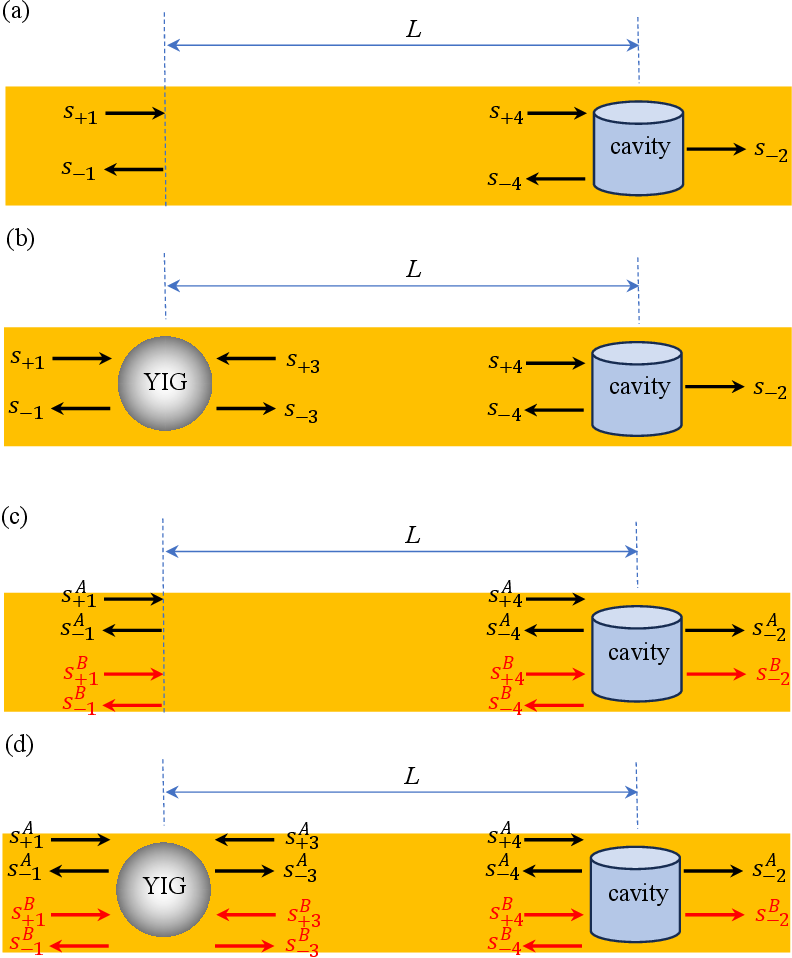}
\caption{ \textbf{Schematic of multi-mode waveguide}. The waveguide (yellow) consists of one or two propagation modes. Each mode can propagate forward and backward. (a) A cavity resonator or (b) a cavity resonator and a YIG sphere are placed on the single-mode waveguide. The positions of YIG and cavity resonator are marked by the vertical dashed lines. $L$ defines the distance between YIG and cavity and is usually a few meters. There are four ports, i.e. 1$\sim$4. At each port, the incoming and outgoing waves are expressed by $s_+$ and $s_-$. In the multi-mode waveguide as shown in (c) and (d), there are more than one propagation modes, see e.g. modes $A$ and $B$. } \label{fig1}
\label{fig1}
\end{figure}

\begin{figure}[t]
\centering
\includegraphics[angle=0,scale=0.5]{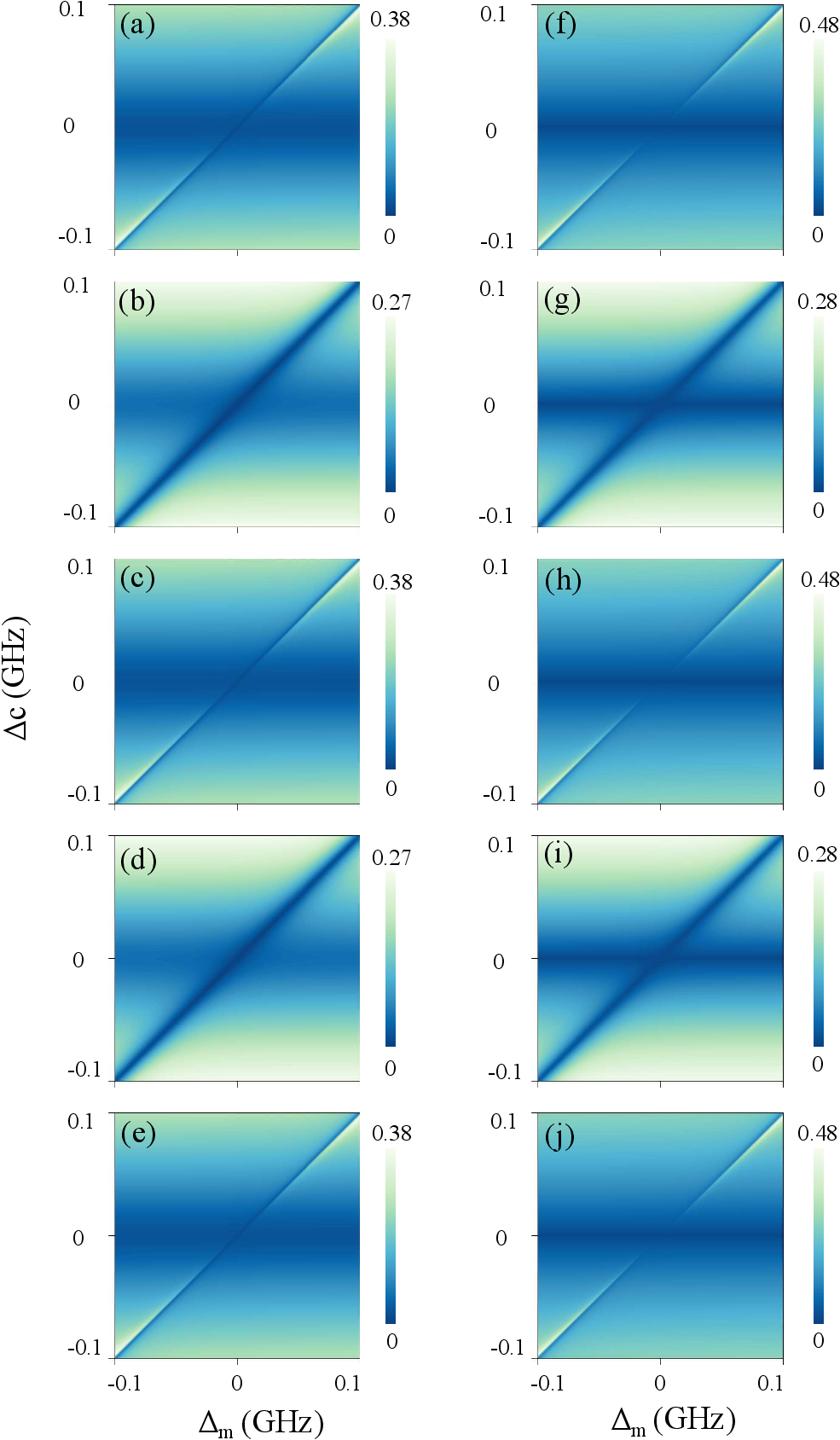}
\caption{ \textbf{Single-mode waveguide}. $|S_{21}|$ spectra as $\beta L$=0 (a), $\pi$/2 (b), $\pi$ (c), 3$\pi$/2 (d) and 2$\pi$ (e) for non-critical coupling condition. (f)$\sim$(j) are for critical coupling condition with $\kappa_{c0}$=0. Here, $\Delta_c$=$\omega-\omega_c$ and $\Delta_m$=$\omega_m-\omega_c$. } \label{fig2}
\label{fig2}
\end{figure}

\begin{figure}[t]
\centering
\includegraphics[angle=0,scale=0.5]{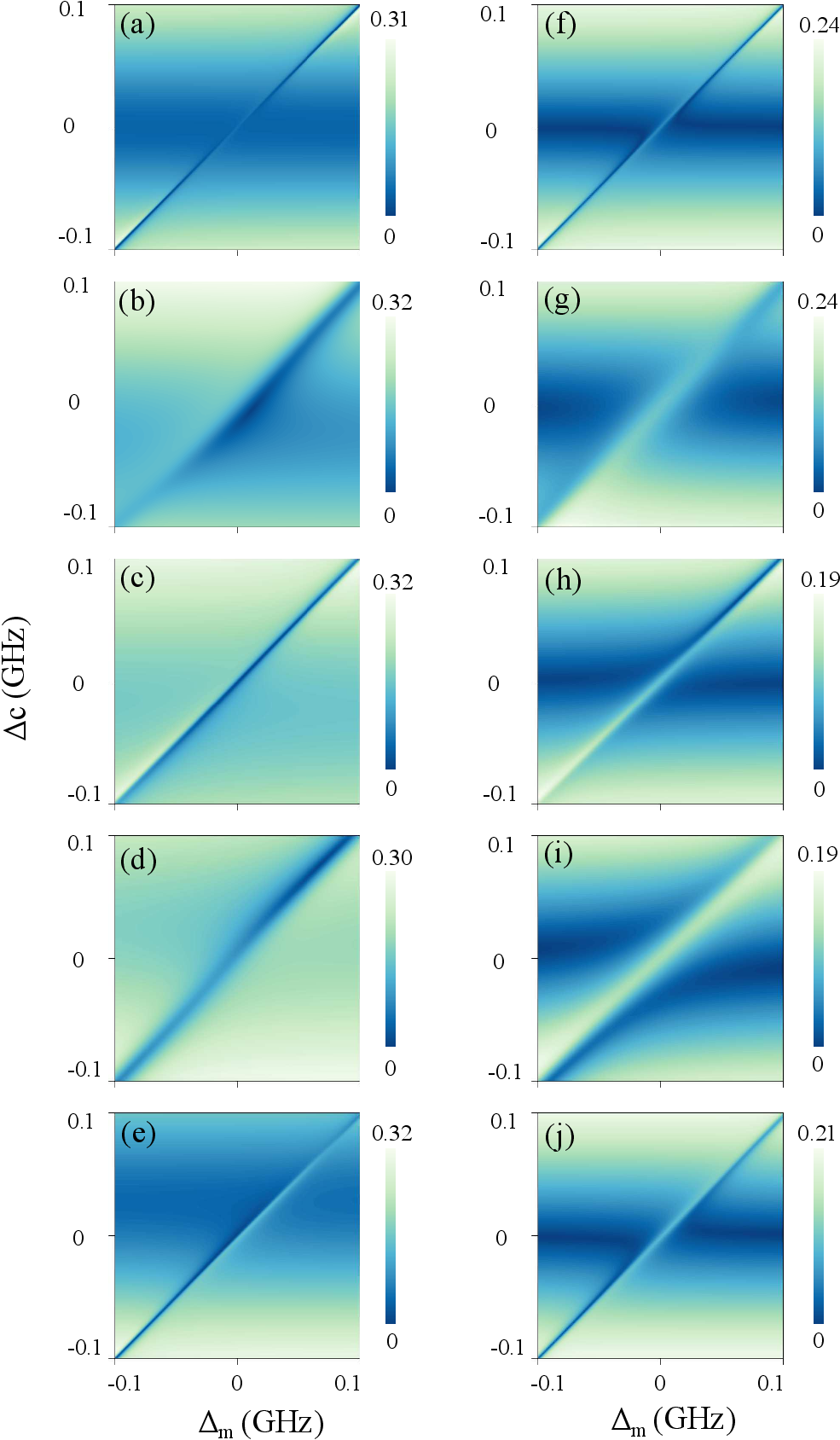}
\caption{ \textbf{Two-mode waveguide}. The same as Fig. 2 but for a two-mode waveguide. Left and right panels show results of non-critical coupling and critical coupling. } \label{fig3}
\label{fig3}
\end{figure}

\section{II. Single-mode waveguide}

Before presenting the results of multi-mode waveguide, we first discuss the results of single-mode waveguide for the sake of comparison. Figure 1 (a) shows the device geometry of a waveguide and a cavity resonator, e.g. a dielectric resonator or a split-ring resonator loaded on a microstrip studied in recent experiment~\cite{yang2024prl}. The length of waveguide is a few meters while the size of resonator is a few millimeters. The incoming microwave is injected from the input port of the waveguide which is marked by the left vertical dashed line. The resonator is placed near the output port of waveguide marked by the right vertical line. The distance between two vertical lines is defined to be $L$. Similar to experimental setup, we assume that $L$ is about several meters which is much larger than the size of resonators. The input port and output port are defined as port 1 and port 2. After traveling with a length of $L$, the incoming microwaves arrive at the port 4. The amplitudes of incoming (outgoing) waves are denoted by $s_{+1}^{}$ ($s_{-1}^{}$), $s_{+4}^{}$ ($s_{-4}^{}$) and $s_{+2}^{}$ ($s_{-2}^{}$) at ports 1, 4 and 2 respectively. Due to the propagation of microwave, a propagation phase will accumulate and thus we have $s_{+4}^{} = e^{-j\beta L} s_{+1}^{}$ and $s_{-1}^{} = e^{-j\beta L} s_{-4}^{}$. $\beta$ is the propagation constant of the mode in the waveguide.

Based on the coupled-mode theory, the dynamical equation of the amplitude of resonator mode $c$ in Fig. 1 (a) is written as~\cite{Manolatou1999ieee}
\begin{align}
\frac{d c}{dt} = \left( j\omega_c - \kappa_{c0} - \kappa_c \right) c + \sqrt{\kappa_4} e^{j\theta_4} s_{+4} \label{single-c},
\end{align}
where $\omega_c$ and $\kappa_{c0}$ are resonance frequency and intrinsic damping of cavity resonator. $\sqrt{\kappa_4}$ and $\theta_4$ is the amplitude and phase of coupling strength between incident microwave and the resonator. Due to the resonator-waveguide coupling, the microwave power could decay from resonator to waveguide and thus an external damping $\kappa_c$ of resonator occurs. As there is no incident wave from output port 2, we have $\kappa_c=\kappa_4$. In this work, we focus on the propagation phase $\beta L$ only and thus neglect all coupling phase $\theta_i$ ($i$=1$\sim$4) at all ports.

Based on the input-output relation~\cite{clerk2010rmp}, the waves at input port 4 and output port 2 satisfy
\begin{align}
s_{-2} = e^{-j\beta d} ( s_{+4} - \sqrt{\kappa_4} c ) \label{inputoutput},
\end{align}
where $d$ is the interaction range of fields of resonator and waveguide. Since $d$ ($\sim$mm) is much smaller than $L$ ($\sim$m), we will drop the phase factor $e^{-j\beta d}$ hereafter. With Eq. (\ref{single-c}) and Eq. (\ref{inputoutput}), the transmission coefficient $S_{21}=\frac{s_{-2}}{s_{+1}}$ is written as
\begin{align}
S_{21} = e^{-j\beta L} \left( 1 - \frac{\kappa_c}{j(\omega-\omega_c) + \kappa_{c0} + \kappa_c} \right) \label{s21-0c}.
\end{align}

Eq. (\ref{s21-0c}) presents two characteristics. First, at resonance ($\omega=\omega_c$), the zero intrinsic damping with $\kappa_{c0}=0$ results in zero transmission coefficient which indicates that the resonator and waveguide are critically coupled. Here, we call $\kappa_{c0}=0$ the critical coupling condition~\cite{yang2024prl}. Second, the factor $e^{-j\beta L}$ provides an overall phase shift of $S_{21}$ without altering the resonance frequency and damping. It implies that changing the position of input port will not affect the transmission spectrum $|S_{21}|$.

We next discuss the magnon-photon coupling mediated by a single-mode waveguide shown in Fig. 1 (b). We place a YIG sphere at the left side of waveguide and apply a static magnetic field to induce a magnetic field-dependent magnon frequency $\omega_m$. The distance $L$ between YIG and microwave cavity resonator is very large so as to prevent direct magnon-photon coupling. The dynamical equations of magnon and cavity photon are written as
\begin{align}
\frac{d m}{dt} &= \left( j\omega_m - \kappa_{m0} - \kappa_m \right) m + \sqrt{\kappa_1} s_{+1} + \sqrt{\kappa_3} s_{+3} \label{single-m1}, \\
\frac{d c}{dt} &= \left( j\omega_c - \kappa_{c0} - \kappa_c \right) c + \sqrt{\kappa_4} s_{+4} \label{single-c1},
\end{align}
where $\kappa_{m0}$ and $\kappa_{m}$ are intrinsic and external damping of magnon. $\sqrt{\kappa_1}$ and $\sqrt{\kappa_3}$ is the coupling strength between magnon and incident waves. The coupling induces damping and thus $\kappa_m=(\kappa_1+\kappa_3)/2$ and $\kappa_c=(\kappa_2+\kappa_4)/2$.

Based on the input-output relation, i.e. $s_{-3} = s_{+1} - \sqrt{\kappa_1} m $ and $s_{-4} = - \sqrt{\kappa_2} c $, and the relation of propagation phase, i.e. $s_{+4}^{} = e^{-j\beta L} s_{-3}^{}$ and $s_{+3}^{} = e^{-j\beta L} s_{-4}^{}$, we arrive at
\begin{widetext}
\begin{equation}
 \frac{d}{dt}
\left( \begin{array}{c}
           m \\
           c \\
\end{array} \right) =
\left( \begin{array}{cc}
           j\omega_m - \kappa_{m0} - \kappa_m \ \ \ \ \ - e^{-j\beta L} \sqrt{\kappa_2\kappa_3} \\
           - e^{-j\beta L} \sqrt{\kappa_1\kappa_4} \ \ \ \ \ j\omega_c - \kappa_{c0} - \kappa_c \\
\end{array} \right)
\left( \begin{array}{c}
           m \\
           c \\
\end{array} \right)
+ \left( \begin{array}{c}
           \sqrt{\kappa_1} \\
           e^{-j\beta L} \sqrt{\kappa_4} \\
\end{array} \right) s_{+1} . \label{single-cm}
\end{equation}
\end{widetext}

By considering the outgoing field with $s_{-2} = s_{+4} - \sqrt{\kappa_4} c $, one can easily obtain $S_{21}$. In numerical calculations, we need to know the values of parameters in the above equations. In Ref.~\cite{yang2024prl}, the experimental parameters are: $\kappa_{m0}$=0.8, $\kappa_{1}$=7, $\kappa_{3}$=8, $\kappa_{c0}$=17, $\kappa_{4}$=370 and $\kappa_{2}$=326. The units of all parameters are MHz. One can see that the coupling strengths of forward and backward propagation for both magnon and cavity photon are different in experiment, indicating an asymmetry in the two opposite directions. In our theory, we focus on the multi-mode effect and neglect such asymmetric effect of coupling strength. The parameters used in our numerical calculations are: $\kappa_{m0}$=1, $\kappa_{1}=\kappa_{3}$=8, $\kappa_{c0}$=17, $\kappa_{4}=\kappa_{2}$=350. In the following, $\kappa_3$ and $\kappa_4$ will be replaced by $\kappa_1$ and $\kappa_2$ respectively.

Figure 2 shows the transmission spectrum $|S_{21}|$ as $\beta L$=0, $\pi$/2, $\pi$, 3$\pi$/2 and 2$\pi$. For non-critical coupling (left panel), no obvious features of coupling, e.g. level repulsion or attraction, can be seen. As for critical coupling (right panel), both magnon and photon can be seen but strong coupling can not be clearly recognized. This shows that the system is in the weak coupling regime, which can be explained by analyzing the magnon-photon cooperativity. From the matrix of Eq. (\ref{single-cm}), one can obtain that the coupling strength of magnon and photon is $e^{-j\beta L} \sqrt{\kappa_1\kappa_2}$. The dampings of two modes are $\kappa_{m0}+\kappa_1$ and $\kappa_{c0}+\kappa_2$ respectively. Therefore, the cooperativity is $\mathcal{C}=e^{-2j\beta L} \frac{\kappa_1\kappa_2 }{(\kappa_{m0}+\kappa_1)(\kappa_{c0}+\kappa_2)}$. Obviously, the amplitude of cooperativity is always less than one and both modes are weakly coupled. Moreover, for both critical and non-critical coupling conditions, the transmission spectrum varies with the propagation phase $\beta L$ periodically with the period of $\pi$, instead of 2$\pi$ observed in experiments~\cite{yang2024prl}. Therefore, single-mode waveguide can not give rise to strong coupling of magnon and photon for both critical and non-critical coupling.

\section{III. Multi-mode waveguide}

For a multi-mode waveguide, many experiments show that one of propagation modes acquires most of injected microwave power while the others share the remaining~\cite{forrer1958jap}. The former and latter are usually called the dominant mode and higher-order mode. In our theory, we consider a dominant mode $A$ and a higher-order mode $B$. At the input port, the incident amplitudes of these two modes are denoted by $s_{+1}^A$ and $s_{+1}^B$. We define the ratio of amplitudes as $\eta=\frac{s_{+1}^B}{s_{+1}^A}$. Since the incident microwave power is mainly distributed in the dominant mode $A$, we choose a small value of $\eta$, i.e. $\eta$=0.1, which indicates that only 1$\%$ of the incident microwave power is loaded in the mode $B$.

 Based on the coupled-mode theory, the dynamical equation of the amplitude of resonator mode $c$ in Fig. 1 (c) is written as
\begin{align}
\frac{d c}{dt} = \left( j\omega_c - \kappa_{c0} - \kappa_c \right) c + \sqrt{\kappa_4^A} s_{+4}^A + \sqrt{\kappa_4^B} s_{+4}^B \label{multi-c},
\end{align}
where $\kappa_4^A$($\kappa_4^B$) is the coupling strength between cavity resonator and propagation mode $A$ ($B$) in the waveguide. Due to the absence of incident wave from output port 2, we have $\kappa_c=\kappa_4^A+\kappa_4^B$. With the input-output relation
\begin{align}
s_{-2}^{A} =  s_{+4}^{A} - \sqrt{\kappa_4^{A}} c  \label{inputoutputA}, \\
s_{-2}^{B} =  s_{+4}^{B} - \sqrt{\kappa_4^{B}} c  \label{inputoutputB},
\end{align}
one can obtain the transmission coefficient $S_{21} = \frac{s_{-2}^{A}+s_{-2}^{B}}{s_{+1}^{A}+s_{+1}^{B}}$.

%With the same procedure as single-mode waveguide, one can obtain the transmission coefficient
%\begin{widetext}
%\begin{align}
%S_{21} = \frac{s_{-2}^{A}+s_{-2}^{B}}{s_{+1}^{A}+s_{+1}^{B}} = \frac{ \left[j(\omega-\omega_c) + \kappa_{c0} + \kappa_c - \kappa_4^A - \sqrt{\kappa_4^A\kappa_4^B} \right] e^{-j\beta_A L} + \left[j(\omega-\omega_c) + \kappa_{c0} + \kappa_c - \kappa_4^B - \sqrt{\kappa_4^A\kappa_4^B} \right] \eta e^{-j\beta_B L} }{(1+\eta) \left[ j(\omega-\omega_c) + \kappa_{c0} + \kappa_c \right]}  \label{s21-multiC},
%\end{align}
%\end{widetext}
% As $\eta$=0 and $\kappa_4^B$=0, i.e. in the absence of mode $B$, Eq. (\ref{s21-multiC}) reduces to Eq. (\ref{s21-0c}) of single-mode waveguide. Moreover, when $\beta_A \neq \beta_B$, one can not extract an overall phase factor as done in the case of single-mode waveguide. This indicates that transmission spectrum $|S_{21}|$ is dependent on the length $L$ or the phase factor $\beta L$.

Since the cavity resonator is critically coupled with the waveguide in experimental measurement~\cite{yang2024prl}, one first derive the critical coupling condition for multi-mode waveguide. By setting $S_{21}$=0 at resonance, we obtain
%\begin{widetext}
\begin{align}
\kappa_{c0} + \kappa_c = \frac{ \kappa_4^A + \sqrt{\kappa_4^A\kappa_4^B} + \left( \kappa_4^B + \sqrt{\kappa_4^A\kappa_4^B} \right) \eta e^{j(\beta_A-\beta_B) L} }{1+\eta e^{j(\beta_A-\beta_B) L}}  \label{critical-multi},
\end{align}
%\end{widetext}
where $\beta_A$ ($\beta_B$) is the propagation constant of mode $A$ ($B$).

When the YIG and magnon are present as shown in Fig. 1 (d), the dynamical equations are written as
\begin{widetext}
\begin{equation}
 \frac{d}{dt}
\left( \begin{array}{c}
           m \\
           c \\
\end{array} \right) =
\left( \begin{array}{cc}
           j\omega_m - \kappa_{m0} - \kappa_m \ \ \ \ \  - \sum\limits_{i=A,B} e^{-j\beta_i L} \sqrt{\kappa_2^i\kappa_3^i} \\
           - \sum\limits_{i=A,B} e^{-j\beta_i L} \sqrt{\kappa_1^i\kappa_4^i} \ \ \ \ \  j\omega_c - \kappa_{c0} - \kappa_c \\
\end{array} \right)
\left( \begin{array}{c}
           m \\
           c \\
\end{array} \right)
+ \sum\limits_{i=A,B}\left( \begin{array}{c}
           \sqrt{\kappa_1^i} \\
           e^{-j\beta_i L} \sqrt{\kappa_4^i} \\
\end{array} \right) s_{+1}^i. \label{multi-cm}
\end{equation}
\end{widetext}
where the external damping are $\kappa_m = \left( \kappa_1^A + \kappa_3^A + \kappa_1^B + \kappa_3^B \right)/2$ and $\kappa_c = \left( \kappa_2^A + \kappa_4^A + \kappa_2^B + \kappa_4^B \right)/2$. For simplicity, we consider symmetric coupling in two opposite directions, i.e. $\kappa_4^{A,B}=\kappa_2^{A,B}$ and $\kappa_3^{A,B}=\kappa_1^{A,B}$ respectively. Therefore, we have $\kappa_m = \kappa_1^A + \kappa_1^B$ and $\kappa_c = \kappa_2^A + \kappa_2^B$.

Using Eq. (\ref{multi-cm}) and the input-output relations, i.e.
\begin{align}
s_{-2}^{A} =  e^{-j\beta_A L} s_{+1}^{A} - e^{-j\beta_A L} \sqrt{\kappa_1^{A}} m - \sqrt{\kappa_4^{A}} c  \label{io-multiA}, \\
s_{-2}^{B} =  e^{-j\beta_B L} s_{+1}^{B} - e^{-j\beta_B L} \sqrt{\kappa_1^{B}} m - \sqrt{\kappa_4^{B}} c  \label{io-multiB},
\end{align}
we can obtain $S_{21}$ in the presence of magnon-photon coupling.

As two propagation modes exist in the waveguide, the determination of parameters becomes complicated. The intrinsic dampings $\kappa_{m0}$ and $\kappa_{c0}$ are unrelated to modes $A$ and $B$ and thus are taken to be values in the treatment of single-mode waveguide. Since most of microwave power is carried by the dominant mode $A$, the parameters of damping and coupling strength for this mode are chosen to be the same as those of the single-mode waveguide in Sec. II. As for the higher-order propagation mode, we make an approximation of $\kappa_{1,2}^B = \eta \kappa_{1,2}^A$. In the theory of waveguide-cavity coupling~\cite{scully2012book,yang2023prb2}, the coupling strength $\kappa_{1,2}$ between waveguide and cavity is dependent on the density of states in the waveguide. The small fraction of total microwave power distributed in the mode $B$ results in small density of states and coupling strength, which justifies the approximation used here.

Another parameter to be determined is the propagation constant $\beta_{A,B}$. The propagation constant is mainly related to the frequency of propagation mode in waveguide. The propagation constant of the dominant mode $A$ is determined by the driving frequency. As for the higher-order propagation mode, we make the second approximation of $\beta_B = \xi \beta_A$ with a small value of $\xi$=0.2 in this work. We choose a small $\xi$ because it induces a slow modulation for the variation of $S_{21}$ as function of the propagation phase $\beta_A L$, whereby reproducing the non-$\pi$ period observed in recent experiment\cite{yang2024prl}.

Figure 3 (a)$\sim$(e) show the the transmission spectrum $|S_{21}|$ of two-mode waveguide under the non-critical coupling condition. They do not present obvious strong coupling features. Therefore, even though the multi-mode effect is introduced, strong magnon-photon coupling can not be achieved in the absence of critical coupling. The situation becomes quite different when the critical coupling condition is satisfied, i.e. Eq. (\ref{critical-multi}) is inserted into Eq. (\ref{multi-cm}) in numerical calculations. The strong coupling can be clearly distinguished for every propagation phase shown in Fig. 3 (f)$\sim$(j). The transmission spectra show LA, LR and then LA at $\beta_A L$=0, $\pi$ and 2$\pi$. By comparing the results to those in Fig. 3 (a)$\sim$(e) and Fig. 2, one can conclude that strong coupling given here arises from the multi-mode waveguide and critical coupling.

 To better understand strong coupling, we first study the effect of critical coupling on the cooperativity $\mathcal{C}$ for multi-mode waveguide. To do so, we obtain the magnon-photon coupling strength and damping rates from the matrix in Eq. (\ref{multi-cm}). Under the critical coupling condition, the cooperativity is simplified as
 \begin{align}
 \mathcal{C} \approx e^{-2j\beta_A L} \cdot \frac{  \kappa_1^A\kappa_2^A + 2e^{j(\beta_A-\beta_B) L} \sqrt{\kappa_1^A\kappa_2^A\kappa_1^B\kappa_2^B} }{ \left(\kappa_1^A+\kappa_1^B \right) \left( \kappa_2^A + \sqrt{\kappa_2^A \kappa_2^B} \right) } \label{C-multi-non1}.
 \end{align}
As $e^{j(\beta_A-\beta_B) L}$=1, we can obtain $|\mathcal{C}| \approx$1. Therefore, only the critical coupling can not give rise to strong coupling.

On the other hand, the two modes $A$ and $B$ in the waveguide offer new pathways of microwave transmission and thereby enhance magnon-photon coupling. To directly see it, we derive an analytical form of transmission coefficient
\begin{widetext}
\begin{align}
S_{21}^{} &\approx e^{-j\beta_A L} - \frac{ \sqrt{\kappa_1^A} \left( e^{-j\beta_A L}\sqrt{\kappa_1^A} + e^{-j\beta_B L}\sqrt{\kappa_1^B} \right) \left( j\omega - j\omega_c - \kappa_{c0} - \kappa_c \right)   }{D} \\
       &- \frac{ \sqrt{\kappa_2^A} e^{-j\beta_A L} \left( \sqrt{\kappa_2^A} + \sqrt{\kappa_2^B} \right) \left( j\omega - j\omega_m - \kappa_{m0} - \kappa_m \right) }{D} \\
       &+ \frac{ \kappa_1^A\kappa_2^Ae^{-j\beta_A L} + \kappa_1^A\sqrt{\kappa_2^A\kappa_2^B}e^{-j\beta_A L} + \sqrt{\kappa_1^A\kappa_2^A\kappa_1^B\kappa_2^B}e^{-j\beta_B L}+ \kappa_2^B\sqrt{\kappa_1^A\kappa_1^B}e^{-j\beta_B L} }{D} \label{proc1} \\
       &+ \frac{ \kappa_1^A\kappa_2^Ae^{-3j\beta_A L} + \sqrt{\kappa_1^A\kappa_2^A\kappa_1^B\kappa_2^B}e^{-j(2\beta_A+\beta_B) L} + \kappa_2^A\sqrt{\kappa_1^A\kappa_1^B}e^{-j(2\beta_A+\beta_B) L} + \kappa_1^B\sqrt{\kappa_2^A\kappa_2^B}e^{-j(\beta_A+2\beta_B) L} }{D}, \label{proc2}
\end{align}
\end{widetext}
where $D$ is the determinant of the matrix in Eq. (\ref{multi-cm}). In the derivation of $S_{21}^{}$, we keep the dominant terms only. The four terms in the numerator of Eq. (\ref{proc1}) represent the output wave contributed by the cavity photon $c$. The first term is the transmission channel from magnon $m$ to cavity $c$ and then output port, which is all mediated by the mode $A$ of waveguide. The second term is similar to the first one but the microwave exits the output port using the mode $B$. The third and four terms represents the transmission channels arising from magnon to cavity using the mode $B$. The waves exit the output port based on the mode $A$ (the third term) and $B$ (the fourth term) of the waveguide. Among these four terms, the first term exists in the case of single-mode waveguide while the remaining three terms occur for the multi-mode waveguide only. These new terms provide extra pathways of coupling magnon and cavity photon and thus enhance the coupling strength. Eq. (\ref{proc2}) is similar to Eq. (\ref{proc1}) but for transmission channels contributed by the magnon $m$ and will not be further explained here.

Compared to single-mode waveguide in Fig. 2, the two-mode waveguide shown in Fig. 3 presents a 2$\pi$-period of transmission as the propagation phase $\beta_A L$ varies, which is in good agreement with experimental observations. Mathematically, the off-diagonal elements of matrix in Eq. (\ref{multi-cm}) consist of two phase factors, i.e. $e^{-j\beta_A L}$ and $e^{-j\beta_B L}$. Since $\beta_B \ll \beta_A$, the higher-order propagation mode $B$ gives rise to a slow oscillation. Since the mode $B$ carries only a small fraction of microwave power, it contributes much less to the transmission than the dominant mode $A$. Therefore, the higher-order mode $B$ induces a slow modulation on the periodic variation of the dominant mode $A$, thereby changing the $\pi$-period to the 2$\pi$-period.

Finally, we discuss possible origin and detection method of higher-order propagation mode. In microstrip line used in experimental measurement~\cite{yang2024prl}, the dominant mode is in general the quasi-TEM mode~\cite{pozar2011book}. In addition to it, the line supports several types of higher-order propagation mode, e.g. TM and TE surface waves. Among them, the TM$_0$ surface mode is a well-studied mode since it has zero cutoff frequency~\cite{pozar2011book}. It exists for any nonzero-thickness dielectrics. Even in some circumstance, the quasi-TEM and TM$_0$ can couple with each other due to the alignment of filed lines of two modes. The other TM$_i$ ($i>$0) higher-order modes have finite cutoff frequency and thus are excited with higher frequency. Since the propagation constant of surface mode is sensitive to the thickness of dielectrics, the transmission would be dependent on the thickness. In future experiments, the transmission measurements of magnon-photon coupling can be performed for microstrip lines with varying thickness. When the line is thin, the propagation constants $\beta$ of surface modes are small~\cite{pozar2011book}. To achieve a 2$\pi$-period in Fig. 3, i.e $\beta L$=2$\pi$, a large YIG/cavity separation $L$ is needed in experiments. However, for a thick microstrip line with large propagation constant, a small $L$ is preferred. By measuring transmission coefficient $S_{21}$ of microstrip line with distinct thicknesses in future experiments, one can further verify the effect of multi-mode waveguide proposed in this work.

\section{IV. Conclusion}

In summary, we have studied long-distance coupling between magnon and photon mediated by a multi-mode waveguide. The multi-mode waveguide presents unique characteristics which do not occur in the single-mode waveguide. First, strong coupling of magnon and photon is achieved in multi-mode waveguide while only weak coupling occurs in single-mode waveguide. Second, as the distance between YIG and cavity resonator varies, a 2$\pi$-period, instead of a $\pi$-period of single-mode waveguide, is obtained in the multi-mode waveguide. These theoretical results are consistent with recent experimental observations. We also analyze possible origin of multiple modes in the waveguide and propose experimental scheme to detect them, which could be validated in future experiments.

\begin{acknowledgements}
\textbf{Acknowledgements.} This work has been funded by the National Natural Science Foundation of China under Grant No. 62374087. The author would like to thank Prof. C. -M. Hu, Prof. K. Xia, and Dr. Y. Yang for many stimulating discussions concerning the possible physics in theoretical and experimental results of long-distance magnon-photon coupling.
\end{acknowledgements}

\textbf{CRediT authorship contribution statement}

\textbf{Yang Xiao}: Writing/review/editing, Writing original draft, Visualization, Validation, Methodology, Investigation, Conceptualization.

\textbf{Declaration of competing interest}

The author declares that he has no known competing financial interests or personal relationships that could have appeared to influence the work reported in this paper.

\textbf{Data availability}

Data will be made available on request.

%\section{APPENDIX: Master equation of dissipative coupling}

%\setcounter{equation}{0}
%\renewcommand\theequation{A\arabic{equation}}

\bibliography{qah-cavity}

\end{document}